\documentclass[conference]{IEEEtran}
\IEEEoverridecommandlockouts
\usepackage{cite}
\usepackage{amsmath,amssymb,amsfonts}
\usepackage{algorithmic}
\usepackage{graphicx}
\usepackage{textcomp}
\usepackage{xcolor}
\usepackage{hyperref}
\def\BibTeX{{\rm B\kern-.05em{\sc i\kern-.025em b}\kern-.08em
		T\kern-.1667em\lower.7ex\hbox{E}\kern-.125emX}}
\begin{document}

\title{Data Matters:\\
	{The Case of Predicting Mobile Cellular Traffic}\
\thanks{The authors gratefully acknowledge the support received from Academy of Finland via the Centre of Excellence in Randomness and Structures, decision number 346308 and the computational resources provided by the Aalto Science-IT project. }
}

\author{\IEEEauthorblockN{Natalia Vesselinova, Matti Harjula and Pauliina Ilmonen}
	\IEEEauthorblockA{Department of Mathematics and Systems Analysis, Aalto University, Finland\\
		\{natalia.vesselinova,  matti.harjula, pauliina.ilmonen\}@aalto.fi}
}


\maketitle

\begin{abstract}

Accurate predictions of base stations' traffic load are essential 
to mobile cellular operators and their users 
as they support the efficient use of network resources
and allow delivery of services that sustain smart cities and roads.
Traditionally, cellular network time-series 
have been considered for this prediction task.
More recently, exogenous factors such as points of 
interest and other environmental knowledge 
have been explored too.
In contrast to incorporating external factors,
we propose to learn the processes underlying 
cellular load generation by employing 
population dynamics data.
In this study, we focus on smart roads and 
use road traffic measures  to 
improve prediction accuracy.
Comprehensive experiments demonstrate that by employing 
road flow and speed, in addition to cellular network metrics,
base station load prediction errors 
can be substantially reduced, by as much as  $56.5\%.$ 
The code, visualizations and extensive results are available on
\url{https://github.com/nvassileva/DataMatters}.
\end{abstract}


\section{Introduction}

Accurate predictions of base stations' (BSs) traffic load are essential 
to mobile cellular operators and their users 
as they support the efficient use of network resources
and allow delivery of services that sustain smart cities and roads.
Significant progress has been made in 
developing deep learning architectures 
for cellular traffic predictions \cite{dl2025ntp}. 
Nonetheless, learning useful representations 
is generally based on 
mobile cellular key performance indicators~(KPIs)
much like it has been done in the past 
with classic statistical models 
for time-series predictions. 
Recent works incorporate exogenous information
namely, auxiliary data about extrinsic factors 
that might have an effect on the cellular load~\cite{jiang2022cellular}.
Typically, these are points of interest
(locations with high user activity). 
 In some of the latest 
 contributions~\cite{gong2024kgda}, \cite{2025sttf},
the contextual description is enriched
by including spatial, temporal and semantic relationships
that describe the BSs and their surroundings.

\setlength{\tabcolsep}{3.1 pt}	
\begin{table}[htbp]
	\caption{
	Employing Population Dynamics in Addition to Cellular Data Improves Prediction Performance by at Least $5.7\%$ and by More Than $50\%$}
	\begin{center}
		\scriptsize{
			\begin{tabular}{|c|rrr|rrr|rrr|rrr|}
				\hline
				\textbf{BS} & \multicolumn{3}{|c|}{\textbf{MAE}} & \multicolumn{3}{|c|}{\textbf{MSE}} & \multicolumn{3}{|c|}{\textbf{MAPE}} & \multicolumn{3}{|c|}{\textbf{RMSE}} \\
				\# & min & mdn & max & min & mdn & max &  min & mdn & max &  min & mdn & max \\
				\hline
				3054051 & 16.0 & 15.0 & 14.0 & 29.7 &29.4 & 29.6 & 27.4 & 23.6 & 27.4 & 16.1 & 16.0 & 16.1  \\
				3086071 & 33.9 & 34.0 & 34.7 & 56.3 & 56.5 & 56.6 & 38.2 & 38.4 & 40.5 & 33.9 & 34.0 & 34.1 \\
				3086081 & 13.9 & 14.1 & 14.4 & 25.1 & 25.2 & 25.4 & 12.5 & 14.1 & 16.3 & 13.5 & 13.5 & 13.6  \\
				317706 & 16.6 & 16.4 & 15.0 & 33.8 & 33.2 & 33.5 & 15.8 & 13.7 & 18.7 & 18.6 & 18.3 & 18.4 \\
				317715 & 8.0 & 8.4 & 7.8 & 14.0 & 13.3 & 11.1 & 15.6 & 15.4 & 20.8 & 7.3 & 6.9 & 5.7  \\
				320280 & 18.3 & 18.4 & 27.4 & 35.3 & 36.8 & 40.0 & 12.5 & 12.5 & 14.3 & 19.6 & 20.5 & 22.5  \\
				320287 & 23.0 & 21.4 &19.5 & 38.9 & 36.8 & 35.1 & 25.5 & 23.5 & 23.4 & 21.8 & 20.5 & 19.4  \\
				3410061 & 12.5 & 12.0 & 12.0 & 20.6 & 20.9 & 20.6 & 30.4 & 31.9 & 33.6 & 10.9 & 11.0 & 10.9 \\
				\hline
				\multicolumn{13}{l}{$^{\mathrm{a}}$Mean Absolute (Percentage) Error (MA(P)E), (Root) Mean Square Error ((R)MSE).}\\
				\multicolumn{13}{l}{$^{\mathrm{b}}$The three values listed per error type are the minimum, median and maximum.}
			\end{tabular}
		}
		\label{tab:improvement}
	\end{center}
\end{table}

In contrast to accounting for external factors,
we propose to learn the intrinsic forces that govern 
the mobile cellular traffic generation.
We hypothesize that
by learning the processes underlying BS load generation,
the accuracy with which cellular volume is predicted 
can be scaled up significantly.
To this end we propose to employ data 
that characterizes population dynamics namely,
the potential sources of cellular load 
and their fluctuations over time 
in addition to cellular KPIs.

We focus on highways and teleoperated and autonomous driving; 
hence, on short-term forecasting, 
which is arguably more challenging 
than mid- and long-term  
but relevant to low-latency and high-reliability services, 
especially those related to smart traffic and vehicles.
We combine road data with 
mobile cellular~KPIs
to enrich the cellular time-series with
population dynamics data.
The computational cost of using road statistics is negligible,
especially compared to recent studies~\cite{gong2024kgda},
which incorporate a large knowledge graph. 
The combined improvement attained from using this graph and 
the developed deep learning architecture in~\cite{gong2024kgda} when contrasted with
state-of-the-art approaches is at most $18\%$~\cite{gong2024kgda}. 
In our study, we record improvements above~$50\%$, Table~\ref{tab:improvement}, 
due to the use of road statistics alone.
The communication overhead between BSs and roadside units, 
which are typically used in smart roads 
where real-time traffic statistics are within reach, 
is of little relevance.

Our primary contributions are:
\begin{itemize}
	\item We are the first to propose and capture 
	the  phenomena intrinsic to
	mobile cellular traffic generation.
	We employ road traffic metrics 
	to gauge the potential sources of BS load
	in highways together with cellular time-series. 
	This leads to substantial improvements in prediction accuracy.

	\item We develop a methodology for 
	generating cellular traffic volumes
	in highways based on real-life road traffic data.
	
	\item We conduct comprehensive experiments 
	for a real-world highway under 
	quickly fluctuating road traffic load, 
	seasonal changes and 
	a diversity of mobile communication volumes, 
	which complement our first study~\cite{vesselinova2023road}.
	
	\item We examine highways
	and address short-term mobile cellular traffic predictions
	in contrast to the extensively studied urban scenario
	and long-term cellular forecasting.
\end{itemize}

The methodology we develop for creating cellular traffic data sets 
based on road traffic is explained in~Section~\ref{sec:methodology}.
A description of the real highway scenario used in the experiments 
can be found in~Section~\ref{sec:highway}. 
The traffic prediction problem 
and the implemented learning model 
are formulated in~Section~\ref{sec:prediction}. 
We discuss the experimental setting, 
methodology and protocol, 
and performance results in~Section~\ref{sec:experiments_results}.
The novelty and impact of the approach 
we propose are highlighted in~Section~\ref{sec:impact}.
The main results and future prospects
are summarized in~Section~\ref{sec:conclusion}.
The code, extensive results and visualizations 
are available on
\url{https://github.com/nvassileva/DataMatters}.

\section{Methodology}
\label{sec:methodology}

Measuring and recording a large variety of different  KPIs
is intrinsic to any communication network because the measurements 
are used for monitoring, operating, optimizing and maintaining the network. 
However, mobile operators rarely share such data because of privacy concerns.
One of the very few open access data sets \cite{amini2023cellular} is 
the Telecom Italia Big Data Challenge~\cite{TelecomItalia2015}.
Therefore, the majority of  the studies, which  develop machine learning methods 
for predicting mobile cellular traffic, routinely use it~\cite{jiang2022cellular}. 
Commonly, their goal  is to predict 
the number of calls as a measure of the cellular network load. 
We adopt the same approach of using the total calls number
as BS load indicator too---we expect  
the general trends we observe in this study 
to remain the same independent of the load traffic measure used.

Similar to the KPIs' relevance to mobile cellular networks, 
road traffic measurements are critical to 
any off-the-shelf traffic management system. 
In contrast to cellular KPIs, however,
traffic measurements are open access.
The California Department of Transportation (Caltrans), for
instance, is maintaining a Performance Measurement System
(PeMS) \cite{chen2001freeway} with real-time traffic data from a
large number of individual detectors deployed statewide in
the California freeway system. Both real-time and historical
measurements are freely available.

We use real-life PeMS data  
to generate mobile cellular traffic
due to the lack of open data sets 
comprising both network KPIs and road traffic measurements.
Before describing our methodology,
we summarize the variables we use from PeMS.

\subsection{Road data}
\label{sec:road_data}
The road metrics employed in this study are
vehicular \textit{flow}---the number of vehicles that pass by a specific
location during a time interval---and \textit{speed} of the flow. 
In the PeMS data set, the former metric is aggregated over 
five-minute intervals and the latter is averaged over these intervals.

We make an important assumption that road traffic is measured outside
yet in the vicinity of the BS of interest, so that
the flow and speed measured in time slot $t$ are observed 
during time slot $t+1$ in the area covered by this BS.

\subsection{Mobile cellular data generation}
\label{sec:cellular_data_method}

We use the described PeMS variables to emulate 
the time of arrival of the vehicles per five-minute time slot,
the dwell time they spent in the cell 
and their departure times. These, together with the 
call duration, are then used to determine if a call 
is handed over to the next BS on the road or
if it is terminated before the vehicle leaves 
its serving BS. 

\subsubsection{Vehicular flow arrival and departure}
The loop detectors of the Caltrans system 
measure all variables in real time yet the measurements are
recorded in PeMS as aggregates for the number of vehicles 
and as averages 
for the speed per five-minute intervals,
as mentioned earlier.  
Vehicular arrival and departures times at a BS
are needed to simulate the call cellular load. 
The classic Poisson assumption about arrival times
has been validated in~\cite{gramaglia}
with real data from~highways 
for sparse---not congested nor experiencing bursts---traffic.
Since for the large majority of the data we employ, the speed is 
around its maximum allowed value, 
suggesting congestion is not habitual,
we adopt this assumption too.
In effect, we emulate each vehicle's arrival time
by exponential inter-arrival times
within each five-minute interval.
The vehicle's departure time is determined by 
its arrival and cell dwell times.

\subsubsection{Dwell time}
The time spent by a vehicle in a segment of a highway served by a BS,
is modeled by the BS's range and the measured PeMS average speed.
We add Gaussian noise with mean zero and standard deviation 0.05 
to emulate different speeds.

\subsubsection{Calls}
New call arrivals at a BS follow a Poisson~process 
as demonstrated empirically, analytically and by simulation~\cite{vesselinova2012admission}.
Therefore, we model the calls placed by each vehicle 
by exponential call inter-arrival times.
Each vehicle generates calls 
with a given rate~$\lambda$,
common for all vehicles.
Although $\lambda$ is fixed, 
the new call arrival process at a BS 
is a non-homogeneous Poisson process,
with time-varying arrival rate 
given by the sum of the time-varying vehicular arrival rate 
(different per five-minute interval) and the fixed call arrival rate.
 As this is a random
process, each passing vehicle might generate no service request,
at most one, or multiple requests while driving through the
segment of the freeway covered by the BS.
Furthermore, during congestion periods,
the simulated load on the network grows correspondingly 
(see~Section~\ref{sec:data}).

We do not model the handover arrival process explicitly.
The number and rate of handovers are determined by 
the start and end times of the simulated calls and the 
vehicle's departure time from a BS. 
Hence, the handover process is completely defined by
the call duration, cell range, vehicle's speed and new call generation process.
If the call is not terminated before the vehicle's departure time,
it is transferred to the next BS. 
Since we do not model the service capacity
of a BS, all calls are accepted by the BS and correspondingly, counted.

\begin{figure}[htbp]
	\centerline{\includegraphics[width=0.47\textwidth]{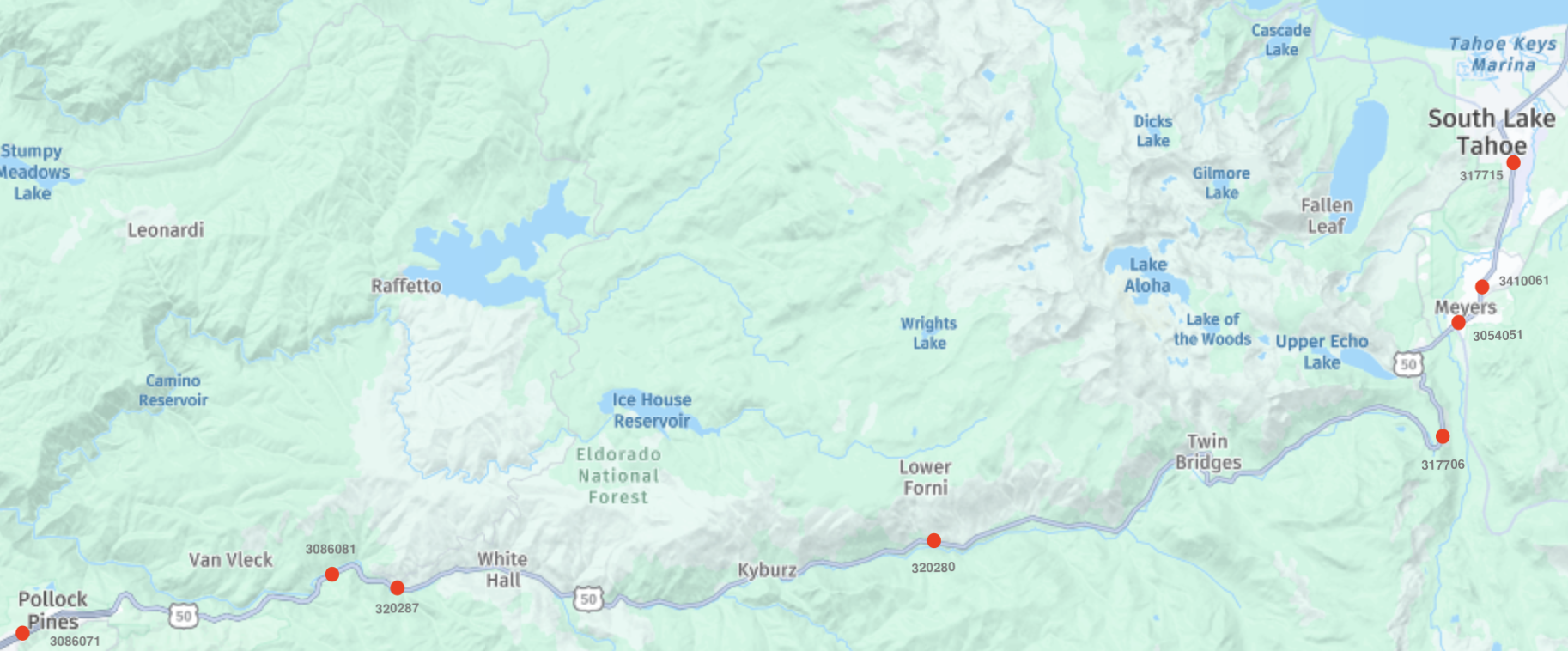}}
	\caption{A map of the section of US50-E El Dorado County freeway used in the study (in gray). The red dots indicate the approximate location of the PeMS detectors on that segment, from Pollock Pines to Lake Tahoe Airport. A larger image is available  on \url{https://github.com/nvassileva/DataMatters}.}
	\label{fig:map}
\end{figure}

\subsubsection{Call duration}
Call duration has been traditionally modeled by the exponential random variable \cite{vesselinova2012admission}.
However, more recent field studies show that a better fit 
for the call duration distribution might be the log-normal distribution \cite{guo2007estimate}
or a mixture of two log-normal distributions \cite{yavuz2007modeling}.
The latter can model diurnal (longer) and nocturnal (shorter) calls  \cite{willkomm2008primary}
or a mixture of two different services, for instance. 

\subsubsection{Flow between adjacent cells}
The vehicular flows in two adjacent BSs can differ from each other.
The recorded number of vehicles in one segment of the highway 
can be smaller or higher than that of the subsequent segment. 
This can be due to arterial roads that merge into the highway.
When the incoming flow in the target cell 
is smaller than the outgoing flow from the preceding cell, 
we randomly remove vehicles together with their corresponding handover calls 
from the generated data set. The goal is to reflect 
the route deviation of the vehicles, namely those that
leave the highway, into the network statistics too. 
When the flow in the target cell is larger than the adjacent, preceding cell, 
no adjustments are made in the handover statistics
generated in the preceding cell.

\section{A highway scenario}
\label{sec:highway}

We consider a sector of the US50-E freeway in El Dorado County, California;
specifically, the one shown in Figure~\ref{fig:map}.
The Caltrans detectors, whose measurements are used in the study,
are listed in Table~\ref{tab:detectors} in sequential order
together with the distance between each two consecutive detectors. 

The density of the PeMS detectors is high 
at the fragments of the highway that 
cross populated areas (such as villages and towns; 
Pollac Pines and Mevers in this particular scenario) and
is much lower on the highway sectors of mountainous regions. 
Long-range BS typically cover rural areas with sparse population, 
whereas short-range BSs provide coverage in densely populated 
(urban) regions. 
Since the density of the road detectors seems to be guided by
similar principles---long-term density of the vehicular traffic---as 
the placement of BSs---population density 
and the particularities of the terrain---we choose 
the range of the BSs to match the distance between sensors. 
This provides us with a variety of diverse use cases. 
The road segments exhibit different capacity, 
average vehicular density and 
propensity for congestion, 
which conditions together with 
the different BSs' ranges
model dynamic, time- and location-varying call volumes. 
For simplicity, we make the assumption that 
the vehicular traffic is unidirectional,
flowing from Pollac Pines to South Lake Tahoe.

\begin{table}[htbp]
	\caption{PeMS US50-E Detectors Used in the Study\\ 
		and the Distance Between Consecutive Ones. \\
		BSs Have the Same ID as the Sensors.}
	\begin{center}
		\begin{tabular}{|l|r|}
			\hline
			\textbf{PeMS detectors} 												& {\textbf{ miles}} \\
			\hline
			Mainline VDS 3086071--50EB JWO Sly Park EB 			& $7.67$\\
			\hline
			Mainline VDS 3086081--50EB at Riverton Barn CCTV 	& $1.62$\\
			\hline
			Mainline VDS 320287--50EB At Ice House 					 & $13.27$\\
			\hline
			Mainline VDS 320280--Wrights Lake Rd						   & $13.41$\\
			\hline
			Mainline VDS 317706--Echo Summit 								  & $3.89$\\
			\hline
			Mainline VDS 3054051--50EB into Luther 50/89 R. & $0.99$\\
			\hline			
			Mainline VDS 3410061--50EB JEO Pioneer Trl 				& $3.22$\\
			\hline
			Mainline VDS 317715--F St 													& 1.46\\
			\hline
			
			\hline
		\end{tabular}
		\label{tab:detectors}
	\end{center}
\end{table}

\section{Mobile Cellular Traffic Prediction: Definition and Solution}
\label{sec:prediction}

We assess the efficacy of our proposed approach 
by training a suitable machine learning model 
with historical cellular data
 \textit{together with} versus \textit{without} 
incorporating road traffic data.
The classic machine learning model we implement 
is individually trained (with the BS's own data) 
for each BS along the highway.

\subsection{Problem Formulation}
\label{sec:formulation}

Mathematically, we denote by $\mathbf{x}^\tau \in \mathcal{R}^p$
a random variable comprising historical measurements of $p$
metrics relevant to a given cell: cellular traffic volume 
or cellular and road traffic volume and speed measurements. 
Given a time sequence of $M$ such historical observations 
$\{ \mathbf{x}^{\tau - M + 1}, \dots, \mathbf{x}^\tau \}$,  
our objective is to learn a prediction model $\mathcal{F}$ 
that can forecast the future mobile traffic load
$\hat{x}$ in the cell during the next time step:

\begin{equation}
	\hat{x}^{\tau + 1} = \mathcal{F} (\mathbf{x}^{\tau - M + 1}, \dots, \mathbf{x}^\tau), \label{eq:focus}
\end{equation}
so that the prediction error is mininized:
\begin{equation*}
	\text{min} \: L(\hat{x}, x),
\end{equation*}
where the loss function $L(\mathord{\cdot})$ measures the difference between 
the estimated $\hat{x}$ and observed $x$ mobile cellular traffic.

\subsection{A classic machine learning model}
\label{sec:model}

Mobile cellular traffic volumes at BSs experience
spatial-temporal dependencies. Since we focus on 
a short-term forecasting problem at a BS \eqref{eq:focus}, 
the impact of correlations is confined to the adjacent BSs.
We account for this phenomenon by the incoming vehicular flow and speed.

\begin{figure*}[htbp]
	\centerline{\includegraphics[width=.45\textwidth]{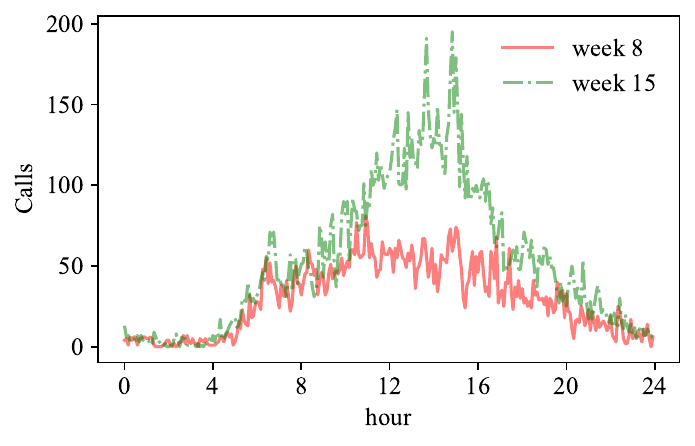}\includegraphics[width=.45\textwidth]{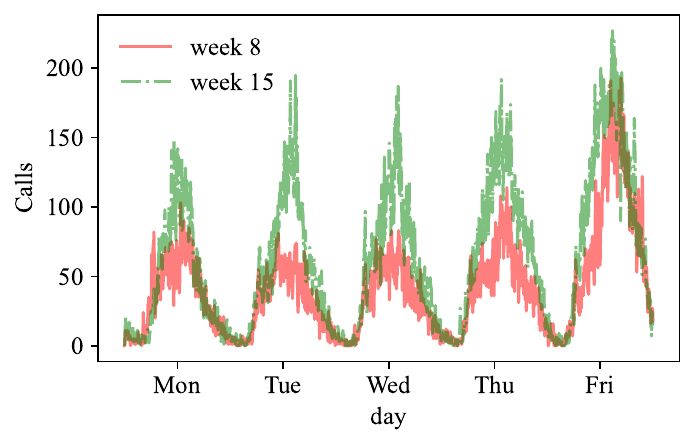}}
	\caption{ BS 320287 Tuesday's and weekly fluctuations of cellular load for two specific weeks.}
	\label{fig:320287}
\end{figure*}

The classic long short-term memory (LSTM) model 
has been broadly applied to time-series prediction tasks 
either as a singular learning structure or as a basic component of 
more complex models.
Furthermore, LSTM has been shown to excel 
in very short-term cellular forecasting~\cite{fang2018mobile},
which motivates our choice.

The model we construct is composed by an LSTM layer, 
followed by 
a fully connected feedforward neural network 
(a multilayer perceptron).
This dense layer consists of a single unit  
and no activation function. 
It is a linear transformation that maps 
the final LSTM state
into a single target value---the estimated number of calls~in~a~BS.

Mathematically, the model can be represented 
as two functions $f(g),$ where
$g(\cdot)$ is the LSTM learning structure,
which transforms the input  data into new features.
The representation function $f(\cdot)$ maps the learned features into 
a cellular traffic load prediction for the BS under consideration.

\section{Experiments and results}
\label{sec:experiments_results}

\subsection{Setting}
\label{sec:setting}

\subsubsection{Input variables' values}
\label{sec:input}

Real road traffic data from Caltrans detectors of the US50-E highway 
in El Dorado County, California are used in the experiments. 
The considered time period is 24-week-long,
from March 28, 2022 to September 9, 2022 
(week 13 to week 36 of 2022).
The data comprise the flow and speed 
from Monday to Friday of each week.
There are 288 data points per day due to the 5-minute granularity.

We generate the network statistics following the  methodology
developed in Section~\ref{sec:methodology} and input values as follows.
New calls are generated with arrival rate $\lambda=1/5,$ one call per 5-minute interval on average.
When defining the mean call duration and its variance, 
we are guided by~\cite{willkomm2008primary} and \cite{guo2007estimate}.
We set the mean call duration 
to 1 and 10~minutes for a mixture of two log-normally distributed calls, 
each with equal weight. 
The variance of the log-normal calls is 3 times larger 
than their mean~\cite{willkomm2008primary}.
The range of the  BSs is listed in Table~\ref{tab:detectors}.

\subsubsection{Data}
\label{sec:data}

All BSs experience a highly dynamic 
road and cellular traffic volumes.
The period from March to September 
accounts for different seasons and 
correspondingly, diverse road traffic patterns  
that reflect seasonal changes and holiday periods.
As a result, each BS has 
different daily and weekly profiles
along the studied period. 
Furthermore, for all BSs
the flow and communication load 
vary quickly within the 5-minute intervals.
Figure~\ref{fig:320287} 
shows the profile of call load 1) for the same location 
and day of a week but different weeks
and 2) for all work days of a week and different weeks
for a BS as an example.

The correlation between 
the road metrics---flow and speed---varies greatly
between segments of the highway and with time: 
strongly inverse, reciprocal 
or lacking linear association.
This variability can be explained by 
the flow-speed correlation dependence on
the capacity of the road segment, 
vehicular flow, terrain, location and time. 

Flow and the generated number of calls 
are strongly correlated. 
This reciprocal linear association, corresponds to 
recent~\cite{primaryuserswhisper} and earlier studies~\cite{becker2013human}, 
which show that there exists a strong correlation 
between population density and mobile cellular use 
in urban scenarios.
Research on population mobility,
in effect, 
relies on such a high correlation 
when using call detail records from mobile operators
to understand and model human mobility
in urban environments~\cite{HelsinkiUni22}.
Furthermore, people's density is often estimated
from mobile phone network data~\cite{HelsinkiUni22}, 
\cite{ricciato2017beyond}. 
Under the assumption of connected vehicles and a highway 
scenario---the setting of our study---the load on the BSs
that serve the highway is exclusively generated by the road traffic
on that highway (services used by the 
autonomous driving vehicles and their passengers). 
Naturally, the highest correlation is between the flow of vehicles
and the new calls they place on their serving BS and correspondingly, 
the total number of calls. 
The correlation between
the flow and number of handovers is often of a smaller magnitude
as not all vehicles have active calls when crossing
the borders between BSs, namely when arriving to a new cell.

The number of calls is in general inversely related to the speed
as whenever the road is congested, the dwell time 
of the vehicle on the road is increased and consequently 
the probability of initiating a new call grows.  
This is in line with the common observation 
that the load on the mobile cellular network increases 
during traffic jams. 

\subsubsection{Machine learning model}
\label{sec:ml_setting}

Our model has a single LSTM layer with 16 cells.
In the training phase, 
we use the root mean square propagation 
and mean squared error (MSE) loss 
for optimizing the models parameters.
We set the length of the historical traffic series 
to six samples (30~min) 
due to the speed with which changes
in the traffic volume occur in 5-minute intervals. 
The prediction horizon is the next 5-minute interval. 
The 24 weeks of data are split 12:6:6 chronologically 
into training, validation, and testing
after which data are shuffled.

\subsection{Evaluation methodology and  protocol}
\label{sec:protocol}

Since our goal is to assess the effectiveness
of employing data intrinsic to
mobile cellular volume generation,
we contrast the prediction performance
of the implemented machine learning model 
when cellular KPIs are used (the baseline) 
with the model's performance 
when these same data
are enriched with vehicular flow and speed metrics.
We evaluate the predictions over
a comprehensive set of highway and cellular conditions.
Similar methodology of assessing 
the impact of data on 
prediction accuracy
is followed in~\cite{gong2023empowering}
when examining the efficacy of 
incorporating an urban knowledge graph.  

We measure the forecasting performance with 
the mean absolute error (MAE), 
mean absolute percentage error (MAPE), 
MSE and root mean squared error (RMSE).
To have a common basis for comparison and 
to fairly attribute the performance deviation
to the data set employed in learning,
we train and evaluate the model 
with the same~hyper-parameters~Section~\ref{sec:ml_setting}
using data from the same period and location
but with different features: 
either containing purely network metrics
or a set of network and road traffic~metrics.
We measure the improvement in prediction 
as the percentage difference in error by
($ErrM_{net} - ErrM_{net\&road}) / ErrM_{net}$,
where $ErrM$ is the error measure.

\setlength{\tabcolsep}{2.3pt}	
\begin{table*}[htbp]
	\caption{Prediction Performance on Calls Dataset and Flow, Speed and Calls Dataset}
	\begin{center}
		\scriptsize{
			\begin{tabular}{|c|ccc|ccc|ccc|ccc|ccc|ccc|ccc|ccc|}
				\hline
				\textbf{ }&\multicolumn{12}{|c|}{\textbf{Calls}} &\multicolumn{12}{|c|}{\textbf{Flow Speed Calls}} \\
				\cline{2-25} 
				\textbf{BS} & \multicolumn{3}{|c|}{\textbf{MAE}} & \multicolumn{3}{|c|}{\textbf{MSE}} & \multicolumn{3}{|c|}{\textbf{MAPE}} & \multicolumn{3}{|c|}{\textbf{RMSE}} & \multicolumn{3}{|c|}{\textbf{MAE}} & \multicolumn{3}{|c|}{\textbf{MSE}} & \multicolumn{3}{|c|}{\textbf{MAPE}} & \multicolumn{3}{|c|}{\textbf{RMSE}} \\
				& min & mdn & max & min & mdn & max &  min & mdn & max &  min & mdn & max &  min & mdn & max & min & mdn & max &  min & mdn & max &  min & mdn & max \\
				\hline
				3054051 & 0.434 & 0.434 & 0.439 & 0.485 & 0.492 & 0.503 & 116.0 & 118.0 & 128.3 & 0.697 & 0.702 & 0.709 & 0.364 & 0.369 & 0.377 & 0.341 & 0.347 & 0.354 & 84.3 & 90.1 & 93.1 & 0.584 & 0.589 & 0.595 \\
				3086071 & 0.282 & 0.283 & 0.287 & 0.154 & 0.155 & 0.157 & 820.8 & 837.2 & 880.6 & 0.392 & 0.393 & 0.396 & 0.187 & 0.187 & 0.187 & 0.067 & 0.067 & 0.068 & 506.8 & 515.6 & 523.9 & 0.259 & 0.259 & 0.261 \\
				3086081 & 0.394 & 0.396 & 0.399 & 0.317 & 0.318 & 0.319 & 335.2 & 347.4 & 365.4 & 0.563 & 0.564 & 0.565 & 0.339 & 0.34 & 0.342 & 0.237 & 0.238 & 0.238 & 293.2 & 298.3 & 305.9 & 0.487 & 0.487 & 0.488 \\
				317706 & 0.155 & 0.156 & 0.158 & 0.058 & 0.059 & 0.060 & 127.5 & 128.4 & 137.8 & 0.241 & 0.242 & 0.244 & 0.129 & 0.13 & 0.135 & 0.038 & 0.039 & 0.040 & 107.4 & 110.8 & 112.0 & 0.196 & 0.198 & 0.199 \\
				317715 & 0.321 & 0.324 & 0.325 & 0.242 & 0.243 & 0.246 & 103.5 & 109.3 & 117.3 & 0.492 & 0.493 & 0.496 & 0.295 & 0.297 &  0.3 & 0.208 & 0.21 & 0.218 & 87.4 & 92.4 & 92.8 & 0.456 & 0.459 & 0.467 \\
				320280 & 0.080 & 0.081 & 0.097 & 0.013 & 0.013 & 0.015 & 54.0 & 54.6 & 57.3 & 0.112 & 0.114 & 0.121 & 0.065 & 0.066 & 0.070 & 0.008 & 0.008 & 0.009 & 47.3 & 47.7 & 49.1 & 0.090 & 0.091 & 0.094 \\
				320287 & 0.066 & 0.067 & 0.067 & 0.009 & 0.009 & 0.009 & 86.7 & 87.0 & 88.3 & 0.094 & 0.094 & 0.094 & 0.051 & 0.052 & 0.054 & 0.005 & 0.006 & 0.006 & 64.6 & 66.6 & 67.7 & 0.073 & 0.075 & 0.076 \\
				3410061 & 0.404 & 0.404 & 0.407 & 0.389 & 0.391 & 0.397 & 696.4 & 725.7 & 757.9 & 0.624 & 0.626 & 0.63 & 0.354 & 0.356 & 0.358 & 0.309 & 0.31 & 0.316 & 484.8 & 494.5 & 503.2 & 0.556 & 0.557 & 0.562 \\
				\hline
				\multicolumn{25}{l}{$^{\mathrm{a}}$The three values listed per error type are the minimum, median and maximum. The MAPE values are in percentage.}
			\end{tabular}
		}
		\label{tab:mixture_fspc}
	\end{center}
\end{table*}

\begin{table*}[htbp]
	\caption{Prediction Performance on New, Handover, and Total Calls Dataset and the Same Dataset but Enriched with Flow and Speed Time-Series}
\begin{center}
	\scriptsize{
		\begin{tabular}{|c|ccc|ccc|ccc|ccc|ccc|ccc|ccc|ccc|}
			\hline
			\textbf{ }&\multicolumn{12}{|c|}{\textbf{New HO Calls}} &\multicolumn{12}{|c|}{\textbf{Flow Speed New HO Calls}} \\
			\cline{2-25} 
			\textbf{BS} & \multicolumn{3}{|c|}{\textbf{MAE}} & \multicolumn{3}{|c|}{\textbf{MSE}} & \multicolumn{3}{|c|}{\textbf{MAPE}} & \multicolumn{3}{|c|}{\textbf{RMSE}} & \multicolumn{3}{|c|}{\textbf{MAE}} & \multicolumn{3}{|c|}{\textbf{MSE}} & \multicolumn{3}{|c|}{\textbf{MAPE}} & \multicolumn{3}{|c|}{\textbf{RMSE}} \\
			& min & mdn & max & min & mdn & max &  min & mdn & max &  min & mdn & max &  min & mdn & max & min & mdn & max &  min & mdn & max &  min & mdn & max \\
			\hline
			3054051 & 0.426 & 0.428 & 0.431 & 0.479 & 0.498 & 0.504 & 101.3 & 103.1 & 114.5 & 0.692 & 0.706 & 0.71 & 0.366 & 0.369 & 0.37 & 0.343 & 0.35 & 0.355 & 83.7 & 88.2 & 91.9 & 0.585 & 0.591 & 0.596 \\
			3086071 & - & - &-  & - & - &-  & - & - &-  & - & - &-  & - & - &-  & - & - &-  & - & - &-  & - & - &-  \\
			3086081 & 0.39 & 0.39 & 0.392 & 0.307 & 0.307 & 0.309 & 326.7 & 337.1 & 349.7 & 0.554 & 0.554 & 0.556 & 0.336 & 0.339 & 0.346 & 0.235 & 0.235 & 0.238 & 294.6 & 300.3 & 321.9 & 0.484 & 0.485 & 0.488 \\
			317706 & 0.154 & 0.155 & 0.158 & 0.056 & 0.057 & 0.06 & 128.1 & 131.7 & 134.0 & 0.237 & 0.24 & 0.245 & 0.131 & 0.131 & 0.133 & 0.039 & 0.04 & 0.041 & 106.6 & 112.0 & 114.1 & 0.199 &  0.2 & 0.202 \\
			317715 & 0.323 & 0.326 & 0.329 & 0.242 & 0.245 & 0.25 & 106.3 & 111.3 & 113.4 & 0.492 & 0.495 &  0.5 & 0.299 & 0.304 & 0.306 & 0.215 & 0.218 & 0.233 & 82.2 & 85.5 & 91.9 & 0.463 & 0.467 & 0.482 \\
			320280 & 0.076 & 0.081 & 0.100 & 0.012 & 0.012 & 0.015 & 54.0 & 55.5 & 56.4 & 0.108 & 0.111 & 0.123 & 0.062 & 0.063 & 0.069 & 0.008 & 0.008 & 0.008 & 44.9 & 46.8 & 48.6 & 0.087 & 0.089 & 0.092 \\
			320287 & 0.066 & 0.066 & 0.067 & 0.009 & 0.009 & 0.009 & 87.8 & 88.7 & 90.6 & 0.093 & 0.094 & 0.094 & 0.052 & 0.053 & 0.055 & 0.005 & 0.006 & 0.006 & 65.5 & 66.0 & 70.4 & 0.073 & 0.075 & 0.078 \\
			3410061 & 0.405 & 0.408 & 0.415 & 0.394 & 0.403 & 0.423 & 590.1 & 639.2 & 655.4 & 0.628 & 0.635 & 0.65 & 0.355 & 0.357 & 0.363 & 0.312 & 0.316 & 0.322 & 458.7 & 474.7 & 483.2 & 0.558 & 0.562 & 0.568 \\
			\hline
			\multicolumn{25}{l}{$^{\mathrm{a}}$BS 3086071 is the first one in our scenario. Therefore, there is no handover traffic from a preceding BS.}
		\end{tabular}
	}
	\label{tab:mixture_fsnhc}
\end{center}
\end{table*}

\begin{table*}[htbp]
\caption{Prediction Performance with Gaussian Noise Added to the Flow} 
\begin{center}
	\scriptsize{
		\begin{tabular}{|c|ccc|ccc|ccc|ccc|ccc|ccc|ccc|ccc|}
			\hline
			\textbf{ }&\multicolumn{12}{|c|}{\textbf{Flow Speed Calls}} &\multicolumn{12}{|c|}{\textbf{Flow Speed New HO Calls}} \\
			\cline{2-25} 
			\textbf{BS} & \multicolumn{3}{|c|}{\textbf{MAE}} & \multicolumn{3}{|c|}{\textbf{MSE}} & \multicolumn{3}{|c|}{\textbf{MAPE}} & \multicolumn{3}{|c|}{\textbf{RMSE}} & \multicolumn{3}{|c|}{\textbf{MAE}} & \multicolumn{3}{|c|}{\textbf{MSE}} & \multicolumn{3}{|c|}{\textbf{MAPE}} & \multicolumn{3}{|c|}{\textbf{RMSE}} \\
			& min & mdn & max & min & mdn & max &  min & mdn & max &  min & mdn & max &  min & mdn & max & min & mdn & max &  min & mdn & max &  min & mdn & max \\
			\hline
			3054051 & 0.369 & 0.371 & 0.375 & 0.348 & 0.354 & 0.358 & 84.0 & 85.9 & 87.8 & 0.59 & 0.595 & 0.598 & 0.367 & 0.368 & 0.372 & 0.343 & 0.348 & 0.354 & 86.3 & 89.4 & 92.7 & 0.586 & 0.589 & 0.595 \\
			3086071 & 0.205 & 0.205 & 0.206 & 0.077 & 0.077 & 0.078 & 594.5 & 599.4 & 602.1 & 0.278 & 0.278 & 0.279 & - & - &-  & - & - &-  & - & - &-  & - & - &-  \\
			3086081 & 0.346 & 0.349 & 0.35 & 0.244 & 0.245 & 0.247 & 293.1 & 315.5 & 320.6 & 0.494 & 0.495 & 0.497 & 0.344 & 0.347 & 0.348 & 0.241 & 0.242 & 0.243 & 311.9 & 314.9 & 315.8 & 0.491 & 0.492 & 0.493 \\
			317706 & 0.134 & 0.136 & 0.138 & 0.04 & 0.042 & 0.043 & 110.0 & 111.3 & 119.5 &  0.2 & 0.205 & 0.207 & 0.135 & 0.135 & 0.138 & 0.041 & 0.042 & 0.043 & 107.5 & 110.8 & 117.1 & 0.202 & 0.204 & 0.207 \\
			317715 & 0.301 & 0.303 & 0.306 & 0.214 & 0.214 & 0.224 & 88.5 & 91.0 & 96.4 & 0.462 & 0.463 & 0.473 & 0.303 & 0.304 & 0.309 & 0.216 & 0.22 & 0.228 & 89.7 & 94.0 & 95.6 & 0.465 & 0.469 & 0.477 \\
			320280 & 0.070 & 0.072 & 0.074 & 0.009 & 0.009 & 0.010 & 50.5 & 50.9 & 51.1 & 0.096 & 0.097 & 0.098 & 0.071 & 0.076 & 0.084 & 0.009 & 0.010 & 0.011 & 49.8 & 50.0 & 52.1 & 0.095 & 0.099 & 0.104 \\
			320287 & 0.057 & 0.057 & 0.058 & 0.006 & 0.006 & 0.006 & 71.7 & 72.9 & 74.8 & 0.079 & 0.080 & 0.080 & 0.056 & 0.057 & 0.058 & 0.006 & 0.006 & 0.007 & 72.9 & 75.0 & 76.4 & 0.078 & 0.079 & 0.083 \\
			3410061 & 0.36 & 0.361 & 0.365 & 0.313 & 0.318 & 0.327 & 481.2 & 495.0 & 514.7 & 0.56 & 0.564 & 0.572 & 0.361 & 0.362 & 0.369 & 0.32 & 0.322 & 0.334 & 475.7 & 496.3 & 510.8 & 0.566 & 0.567 & 0.578 \\
			\hline
			\multicolumn{25}{l}{$^{\mathrm{a}}$BS 3086071 is the first one in our scenario. Therefore, there is no handover traffic from a preceding BS.}
		\end{tabular}
	}
	\label{tab:noise}
\end{center}
\end{table*}

\begin{figure*}[htbp]
	\centerline{\includegraphics[width=.45\textwidth]{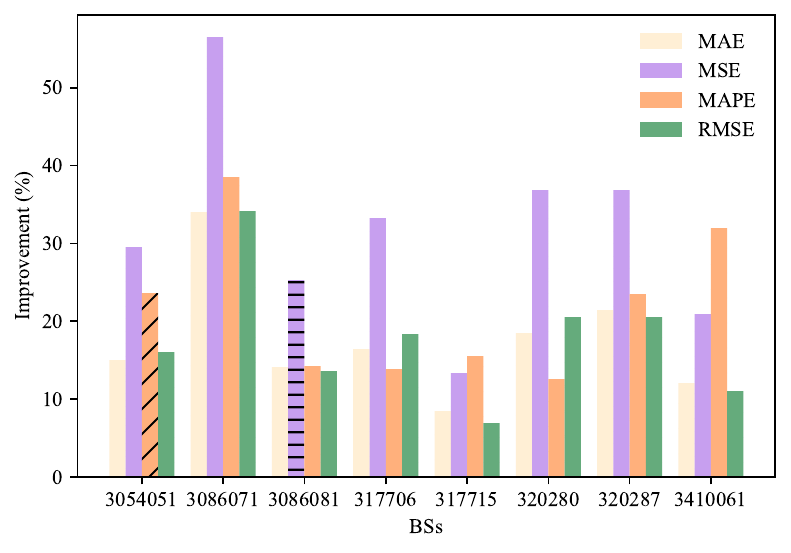} \includegraphics[width=0.45\textwidth]{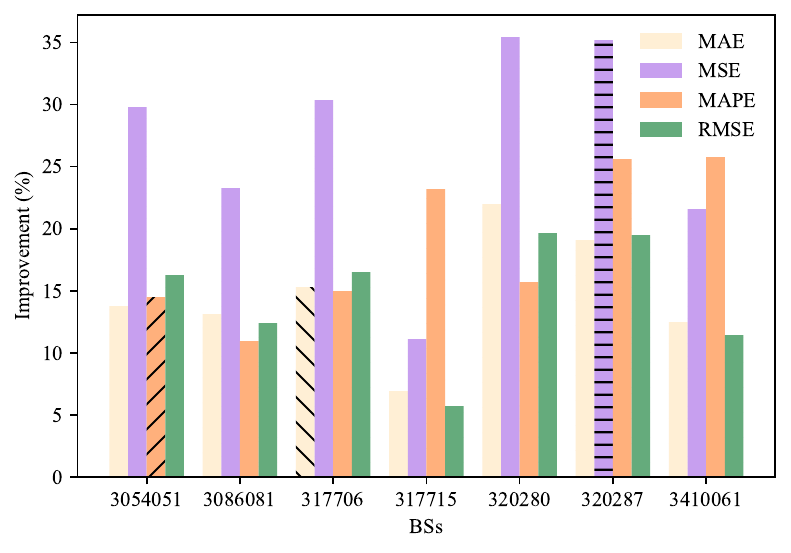}}
	\caption{Improvement in prediction when employing (\textbf{left}) calls vs calls, flow and speed data and (\textbf{right})  new, handover and total number of calls vs the same data set enriched with flow and speed statistics.}
	\label{fig:improvement}
\end{figure*}

\begin{figure*}[htbp]
	\centerline{\includegraphics[width=.35\textwidth]{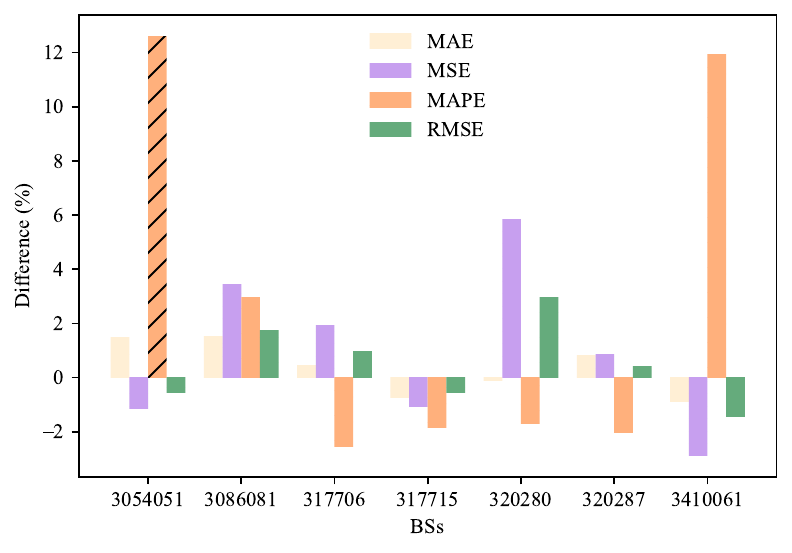} \includegraphics[width=.35\textwidth]{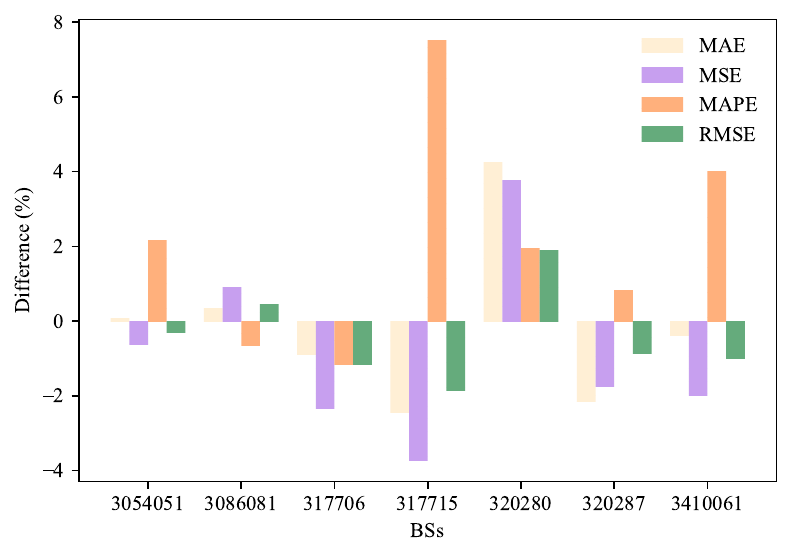} \includegraphics[width=.35\textwidth]{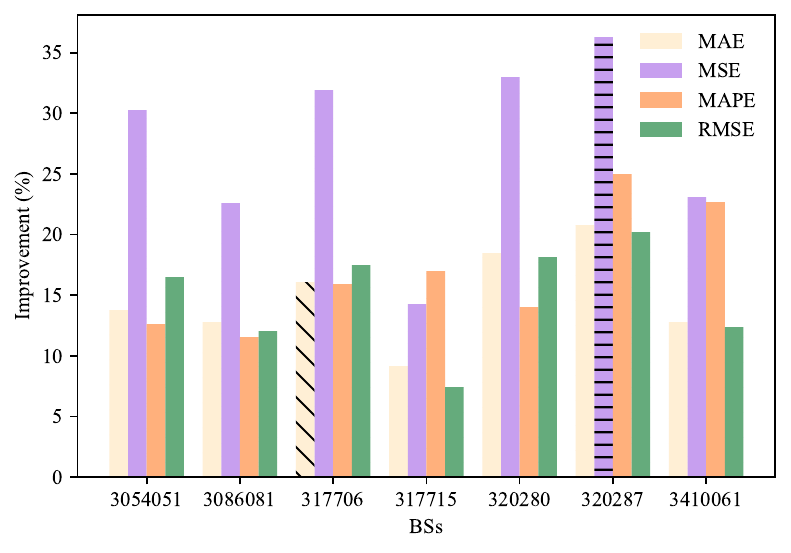}}
	\caption{Prediction error reduction when contrasting: (\textbf{left}) calls (baseline) vs the same data (C) augmented with new,  handover and calls (NHC); (\textbf{center})~flow, speed, calls (FSC)~(baseline) vs flow, speed, new, handover and total number of calls (FSNHC). (\textbf{right}) NHC data set  (baseline) vs FSC.}
	\label{fig:handover_effect}
\end{figure*}

\begin{figure*}[htbp]
	\centerline{\includegraphics[width=.35\textwidth]{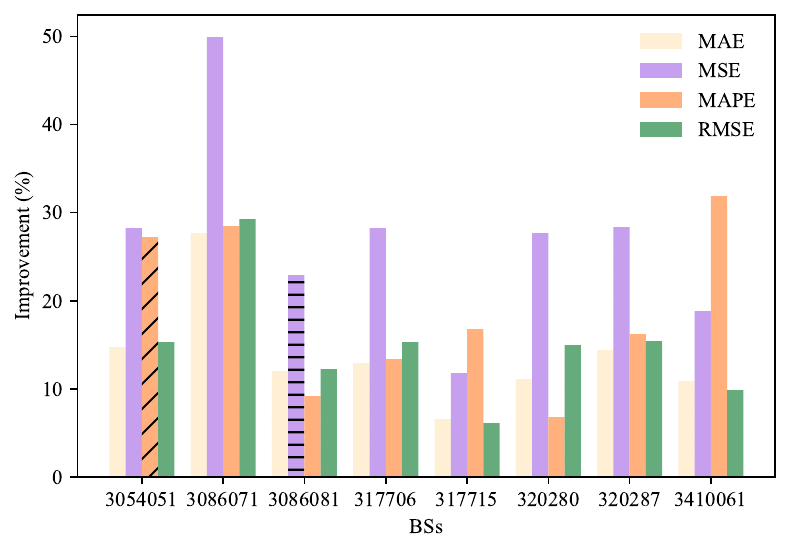} \includegraphics[width=.35\textwidth]{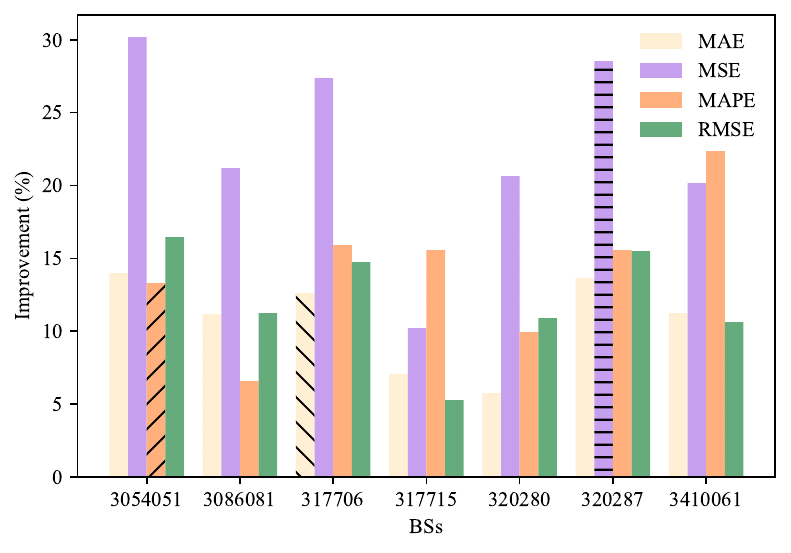} \includegraphics[width=.35\textwidth]{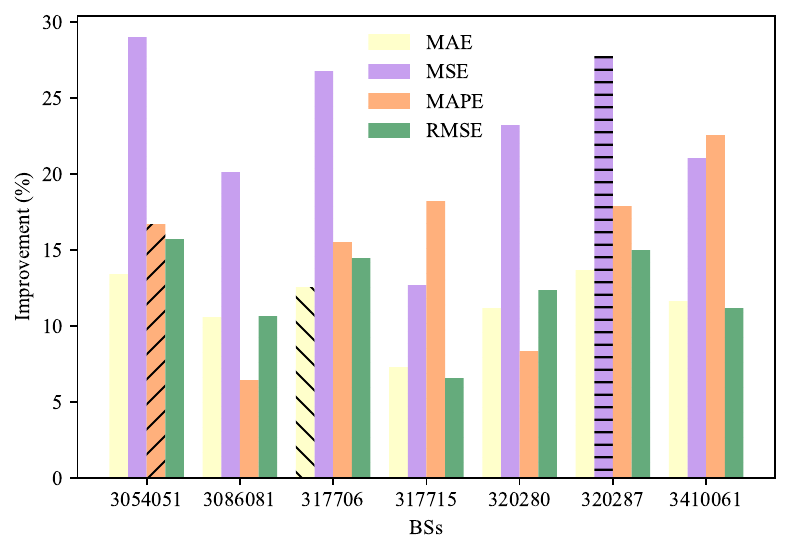}}
	\caption{Improvement in prediction when the flow is estimated with errors. Contrasting: (\textbf{\textbf{left}}) calls (baseline) vs  the same data (C) augmented with vehicular flow estimated with errors and speed ($\hat{F}$SC);
		(\textbf{center}) NHC (baseline) vs $\hat{F}$SNHC. (\textbf{right})~NHC (baseline)~vs~$\hat{F}$SC.}
	\label{fig:noise_effect}
\end{figure*}

\subsection{How much can population dynamics data help?}
\label{sec:performance}

\subsubsection{Overall performance}
Comprehensive results from the four different data sets---1)~calls 
(denoting total number of calls, that is, the sum of new and handover calls) only; 
2)~flow, speed and calls;
3)~new, handover, calls 
and 4)~flow, speed, new, handover, 
calls---when the call duration is modeled by 
a mixture of two log-normal distributions
are summarized in Table~\ref{tab:mixture_fspc}
and Table~\ref{tab:mixture_fsnhc}.
For each error metric we show the minimum, median and maximum
of the 5 simulation runs conducted for each data set and BS.
We make the observations as follows:
\begin{itemize}
	\item \textbf{Employing data that captures the processes
		intrinsic to mobile cellular traffic generation
		consistently improves cellular load prediction performance.}
	All BSs observe improvements in all their forecasting measures 
	when learning from both network and road traffic data.
	
	The prediction error reduction among all BSs
	is between $8.4\%$ and $33.9\%$~(MAE),
	$13.3\%$ and $56.5\%$~(MSE), $12.5\%$ and $38.4\%$~(MAPE), and 
	$6.9\%$ and $34\%$~(RMSE) when in addition to (total number of) calls, 
	also flow and speed are employed, and when comparing the medians of all BSs, 
	see Figure~\ref{fig:improvement} (left). 
	Similar improvement trends are recorded 
	for the minimum and maximum errors,
	Table~\ref{tab:improvement}. 
	Using more refined network KPIs---those that contain the number of new and handover calls,
	and their total---and road data, reduces the prediction errors too: 
	between $7.5\%$ and $21.8\%$~(MAE),
	$11\%$ and $35.4\%$~(MSE), $10.9\%$ and $25.7\%$~(MAPE), 
	and $5.7\%$ and $19.7\%$~(RMSE) for the medians,  
	see Figure~\ref{fig:improvement} (right), and likewise for the smallest and largest errors.
	
	\item \textbf{Road traffic metrics can capture
		the processes underlying  
		mobile cellular load generation.}
	We employ flow and speed as a means to characterize 
	the underlying population dynamics in highways 
	and through them the data generation process. 
	The flow captures the number of potential call generation sources. 
	The speed can indicate density dynamics and
	can also serve as a gauge for road traffic congestion 
	and hence, for increased load on the BS. 
	The number of handovers (or equivalently, the handover rate) 
	is an indicator of mobility too yet our results are not conclusive about its impact. 
	Contrasting the performance of calls to that of new, handover, and calls,
	we observe an error improvement in some of the error measures: 
	it is in the order of no more than $3\%$ for the majority of the BSs, 
	see Figure~\ref{fig:handover_effect}~(left and center plots).
	
	Finally, employing flow, speed and calls (3 variables) leads to a large reduction in the 
	prediction errors even when contrasted with employing new, handover and calls (3 variables):
	$8.6\%$ and $22.8\%$~(MAE),
	$14.2\%$ and $36.3\%$~(MSE), $11.5\%$ and $25\%$~(MAPE), and 
	$7\%$ and $20.2\%$~(RMSE) for the medians, Figure~\ref{fig:handover_effect} (right plot).
	
	\item \textbf{Choice of data (split) and hyper-parameter tuning.}
	The shown results are from using 24~weeks of data 
	and 75:25 ratio of train and validation to test data. 
	We observed cases when 
	the performance of some of the BSs 
	is decreased when using road data. 
	This is the case for the MSE and RMSE of BS~320287 when 
	the data set consists of weeks 15 to 21 and the data is chronologically split~
	into 3:2:2 for training, validation and testing\footnote{See
	\url{https://github.com/nvassileva/DataMatters} for detailed results.}.
	We identify week 15 as outlying 
	from the other training weeks and consequently,
	as the main explanation for the results.
	For the same period but with 4:1:2 partition of the data, the results 
	are consistent with what we report above---decreased prediction errors
	across all performance measures and BSs 
	when employing flow and speed time-series
	in addition to call time-series data.
	This 4:1:2 data split also turns into smaller forecasting errors for the purely network data sets.
	In other words, 
	choice of data that captures the underlying patterns 
	and an appropriate hyper-parameter tuning are always required. 
	In our on-going work we are studying 
	the effect of the data split on the overall forecasting performance. 
\end{itemize}

\subsubsection{Sensitivity to road metrics accuracy}

We envision that road traffic measurements can directly be fed into the learning algorithms
as smart cities and roads are equipped with detectors that can provide such critical information.
Another alternative is to use road traffic predictions instead of measurements. 
The machine learning research community has seen a surge of deep learning methods that tackle this 
prediction task with several solutions that achieve high levels of forecasting accuracy~\cite{yin2021deep}.\footnote{
An advantage of such an approach is that the mobile cellular prediction algorithms can receive
in advance information about the vehicular flow and speed that will be evidenced in the BS during the next time slot. 
This timely received data can increase further the prediction accuracy of the algorithms. 
In fact, our preliminary results validate this concept under similar conditions as those we report above.}

To assess the sensitivity of the mobile cellular predictions to estimation errors in the road predictions,
we introduce an estimation error in the flow measurements by the PeMS detectors. 
In particular, we assume a prediction error of $5\%$ in the number of vehicles per 5-minute intervals, 
modeled by adding Gaussian noise to the real PeMS measurements:
\begin{equation}
	\hat{v} =  v + \epsilon, \qquad \epsilon \sim \mathcal{N}(0,\sigma), 
	\label{eq:est_error}
\end{equation}
where $v$ denotes the real PeMS flow data and $\hat{v}$  is the estimated flow.
Whereas our learning model is fed with the estimated value of the flow variable $\hat{v}$,
we generated the mobile cellular traffic load with the true PeMS measurement $v$. 

The results are reported in Table~\ref{tab:noise} and show that 
although the noise in the flow variable decreases cell load prediction accuracy,
employing road data is largely beneficial yet. Overall, across all error measures 
and all BSs and the 4 data sets, the prediction error improvement is 
between $5.2\%$ and $49.9\%$,  
Figure~\ref{fig:noise_effect}~(left and center plots), 
for the medians when using estimated (namely, with errors) 
road traffic time-series data. 

Employing flow, speed and calls---when flow is reported with errors---leads to a reduction in the 
prediction errors even when contrasted with new, handover and calls data set: the improvement
is between $6.4\%$ and $29\%$ for the medians among the MAE, MAPE, MSE, and RMSE error measures,
Figure~\ref{fig:noise_effect} (right plot).

\section{Novelty and impact}
\label{sec:impact}

\textbf{Novelty}. Mobile cellular load forecasting is native to network 
resource optimization and delivery of services 
with reliability, latency and quality guarantees. 
Therefore, it has remained prominent 
across all generations of communication networks.
In the latest research in this area,
the primary focus is on developing 
larger and more powerful machine learning models~\cite{dl2025ntp}, \cite{jiang2022cellular}.
When it comes to data, the state-of-the-art contributions
(see \cite{jiang2022cellular} for an overview)
focus on incorporating exogenous information.
In particular, these studies rely on 
environmental data---in addition to
cellular time-series---as such auxiliary information
describes the BS context and 
can bring improvements in terms of reduced prediction errors~\cite{gong2024kgda}, \cite{2025sttf}.
A few deep learning algorithms~\cite{fang2022sdgnet}, \cite{handover20}
explicitly incorporate handover rates between BSs
in an attempt to better reflect traffic fluctuations via user mobility.
Nonetheless, none of the prior art proposals 
explores data that can implicitly or explicitly
characterize the data sources and their variability.
The novelty of our approach consists of modeling the internals of 
the cellular traffic generation phenomenon. 
We do this by counting the potential sources of  cellular traffic
and by accounting for their velocity. 

\textbf{Impact.} Employing population dynamics
has the potential not only to reduce the uncertainty 
about future cellular volumes but also to
reduce the cost of collecting big and diverse
data and hence, to reduce the computational complexity 
and memory requirements and ultimately, to decrease the energy expenditure too.
Lighter learning models are also more versatile
as they can be used at the edge and in roadside units too. 
Furthermore, concept drift~\cite{dl2025ntp} arising 
from changes in network configurations or conditions and users' behavior
would have a major impact on models relying on knowledge graphs 
as these graphs would need to be updated first. 

\textbf{Potential limitations.} The concept of 
employing population dynamics
together with relevant mobile cellular data
to accurately predict cellular traffic
is of a much broader scope than a specific scenario. 
We validate our idea in a highway scenario 
yet we conjecture
that it is applicable to the urban setting too.
At present, a limitation for the implementation of our idea in practice 
is the potential lack of real-time population density and mobility data.
Currently, real-time road traffic measurements are 
an integral part of the urban transportation management systems 
and these can be used in forecasting the 
service load placed by vehicles and their passengers
on the cellular network. 
In the future, different means for estimating pedestrian density
will be available too. 
The application of optical cables for simultaneous 
data transmission and tracking of people in a train station,
for instance, is discussed in~\cite{singh2022simultaneous}.

\section{Conclusion}
\label{sec:conclusion}

Mobile cellular data have been used to model population dynamics
with the aim to better understand urban mobility.
We take a different look 
and employ population dynamics in a highway  
to forecast mobile cellular traffic on the highway. 
Our contributions stem from proposing 
the type of data that can, in practice, 
substantially scale up the accuracy 
of mobile cellular traffic predictions.
In particular, we propose to employ data that 
characterizes the sources of communication volume
and its dynamics in addition to BS traffic load.

We validate our hypothesis in a real highway scenario 
by using vehicular flow and speed time-series.
The results show that 
the traffic load prediction error 
at each BS
can be largely and consistently reduced: 
between about $5\%$ and $56.5\%$ depending on 
the error measure, BS and specific data sets. 
These results are obtained for BSs 
featuring realistic conditions 
including different range,
fluctuating road traffic and cellular volumes.

We sincerely hope that this powerful result
will prompt future research.
Our proposal can boost 
the prediction accuracy of 
centralized and distributed learning models 
for short-, mid- and long-term cellular load forecasting.
The idea of using population dynamics
together with relevant mobile cellular data,
is of much broader scope than a specific scenario. 
Examining and quantifying the reductions in prediction error 
with mobile operator and road data on highways and urban scenarios would be the next step.

\bibliography{references}

\end{document}